\documentclass[fleqn,twoside]{article}
\usepackage{espcrc2}

\usepackage{graphicx}

\newcommand{\AmS}{{\protect\the\textfont2
  A\kern-.1667em\lower.5ex\hbox{M}\kern-.125emS}}

\newcommand{\sba}{\mbox{$\sin ^2 (\beta -\alpha)$}}
\newcommand{\mH}{\mbox{$m_{\mathrm{H}}$}}
\newcommand{\mZ}{\mbox{$m_{\mathrm{Z}}$}}
\newcommand{\bb}{\mbox{$\mathrm{b} \bar{\mathrm{b}}$}}
\newcommand{\mh}{\mbox{$m_{\mathrm{h}}$}}
\newcommand{\mA}{\mbox{$m_{\mathrm{A}}$}}
\newcommand{\ee}{\mbox{$\mathrm{e}^{+}\mathrm{e}^{-}$}}
\newcommand{\ra}{\mbox{$\rightarrow$}}

\title{MSSM Constraints from Higgs Boson Searches}

\author{Andr\'e Sopczak, Lancaster University}

\begin{document}
\begin{titlepage}
\def\thefootnote{\fnsymbol{footnote}}       

\begin{center}
\mbox{ } 

\end{center}
\vskip -1.0cm
\begin{flushright}
\Large
\vspace*{-2cm}
\mbox{\hspace{10.2cm} hep-ph/0112086} \\
\mbox{\hspace{12.0cm} December 2001}
\end{flushright}
\begin{center}
\vskip 3.0cm
{\Huge\bf
MSSM Constraints 
\smallskip
from Higgs Boson Searches}
\vskip 1cm
{\LARGE\bf Andr\'e Sopczak}\\
\smallskip
\Large Lancaster University

\vskip 2.5cm
\centerline{\Large \bf Abstract}
\end{center}

\vskip 3.cm
\hspace*{-1cm}
\begin{picture}(0.001,0.001)(0,0)
\put(,0){
\begin{minipage}{16cm}
\Large
\renewcommand{\baselinestretch} {1.2}
The LEP era has brought immense progress in searches for Higgs
bosons over the last 12 years which will guide searches
at future colliders.
The evolution of the Higgs boson mass limits is reviewed with the 
focus on results from general parameter scans in the Minimal Supersymmetric
extension of the Standard Model (MSSM) in contrast 
to the so-called benchmark limits.
The hint for a Standard Model (SM) Higgs boson of 115.6~GeV can also be 
interpreted as a preference for a Higgs boson of that mass in the MSSM.
Further small data excesses allow the hypothesis that
the neutral Higgs bosons of the MSSM all have masses between 
90 and 116~GeV.
\renewcommand{\baselinestretch} {1.}

\normalsize
\vspace{3cm}
\begin{center}
{\sl \large
Presented at the 
Seventh Topical Seminar on the Legacy of LEP and SLC, 
Siena, Italy, October 2001
\vspace{-3cm}
}
\end{center}
\end{minipage}
}
\end{picture}
\vfill

\end{titlepage}


\newpage
\thispagestyle{empty}
\mbox{ }
\newpage
\setcounter{page}{1}

\begin{abstract}
The LEP era has brought immense progress in searches for Higgs
bosons over the last 12 years which will guide searches
at future colliders.
The evolution of the Higgs boson mass limits is reviewed with the 
focus on results from general parameter scans in the Minimal Supersymmetric
extension of the Standard Model (MSSM) in contrast 
to the so-called benchmark limits.
The hint for a Standard Model (SM) Higgs boson of 115.6~GeV can also be 
interpreted as a preference for a Higgs boson of that mass in the MSSM.
Further small data excesses allow the hypothesis that
the neutral Higgs bosons of the MSSM all have masses between 
90 and 116~GeV.
\vspace{1pc}
\vspace*{-1cm}
\end{abstract}

\maketitle

\section{Introduction}
\vspace*{-1mm}
The search for Higgs bosons has been one of the most 
important lines of research during the LEP era.
Final LEP-1 results from data taken
at the Z-resonance~\cite{physrep} and preliminary LEP-2
results for complete data up to 209~GeV center-of-mass 
energy~\cite{moscow}
were recently reviewed.
The MSSM is well motivated and the most discussed extension
of the SM. It predicts three neutral\,(h,\,H,\,and~A) 
and two charged Higgs bosons.
\mbox{ At the beginning} of LEP operation in 1989,
no radiative corrections to the neutral Higgs boson masses
were calculated and $\mh<\mZ$ was predicted. 
\mbox{First- and} second-order~corrections~increased the 
upper mass bound to about 130~GeV. 
The neutral Higgs boson masses and production cross sections
strongly depend on various model parameters. 
At LEP two ways were chosen to present the mass 
limits: benchmark limits for a certain set of MSSM 
parameters and mass limits based on MSSM parameter scans.
Details of the benchmark and scan parameters are given
for example in Refs.~\cite{mssmlep209,209s}.

\section{Largely reduced h mass limit}
\begin{figure}[t]
\vspace*{-0.5cm}
\includegraphics[width=0.43\textwidth]{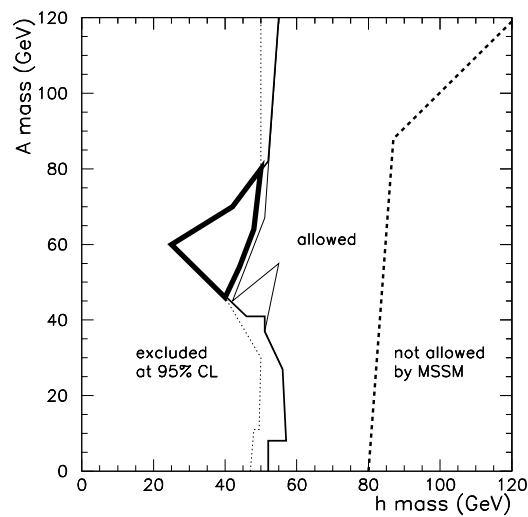}
\vspace*{-1cm}
\caption{91~GeV $(m_{\rm h},m_{\rm A})$ result.}
\label{fig:lep1}
\vspace*{-1cm}
\end{figure}

\vspace*{-1mm}
The importance of a MSSM parameter scan to set mass
limits was already realized at LEP-1. The region marked 
by the thick black line in Fig.~\ref{fig:lep1} shows that
the h mass limit is reduced from 41 to 25 GeV
in a parameter scan compared to benchmark results~\cite{91b,91s}.

\section{Removal of A mass limit}
\begin{figure}[b]
\includegraphics[width=0.40\textwidth]{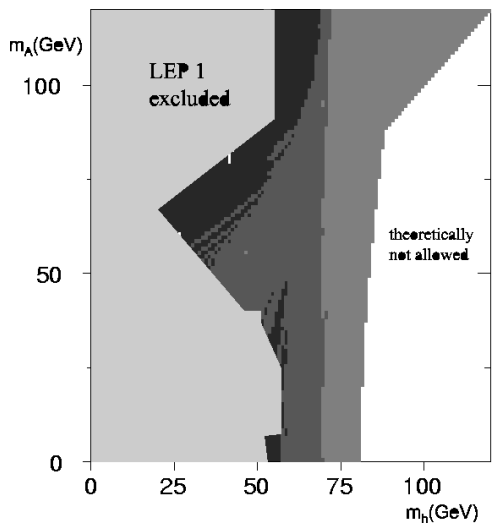}
\vspace*{-0.9cm}
\caption{172~GeV $(m_{\rm h},m_{\rm A})$ result. 
         Only the dark region is excluded in addition to LEP-1.}
\label{fig:172ha}
\vspace*{-1.0cm}
\end{figure}

\vspace*{-1mm}
An early~LEP-2~study showed that a benchmark 
limit of 52 GeV on the A mass~\cite{172b} disappeared 
completely in a MSSM parameter scan~\cite{172s} (Fig.~\ref{fig:172ha}).
\vspace*{-0.5cm}

\clearpage
\section{Similar\,limits\,from\,different\,experiments}
\begin{figure}[h]
\vspace*{-3cm}
\includegraphics[width=0.47\textwidth]{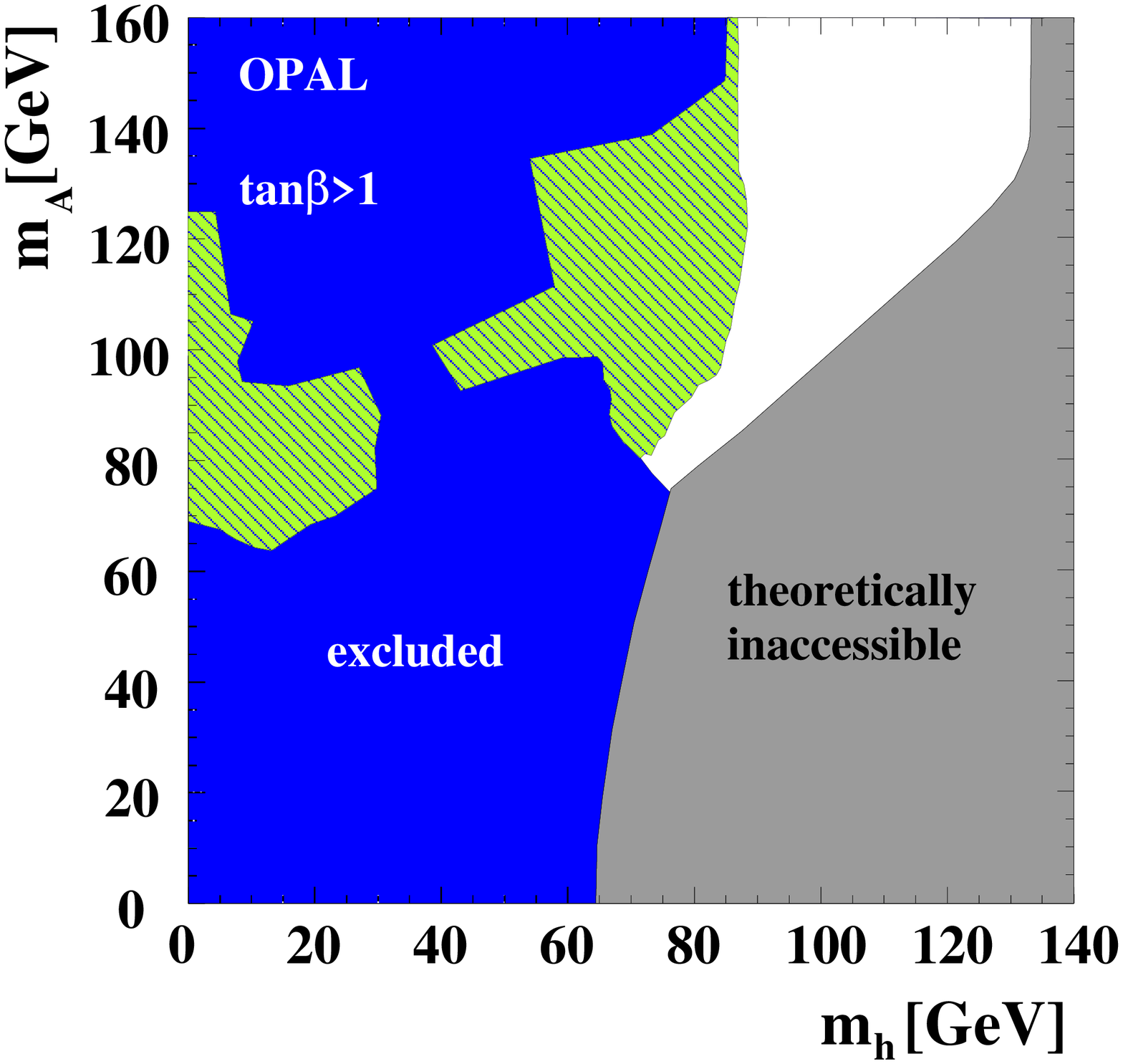}
\vspace*{-1.5cm}
\caption{189~GeV $(m_{\rm h},m_{\rm A})$ result.}
\label{fig:189opalha}
\includegraphics[width=0.47\textwidth]{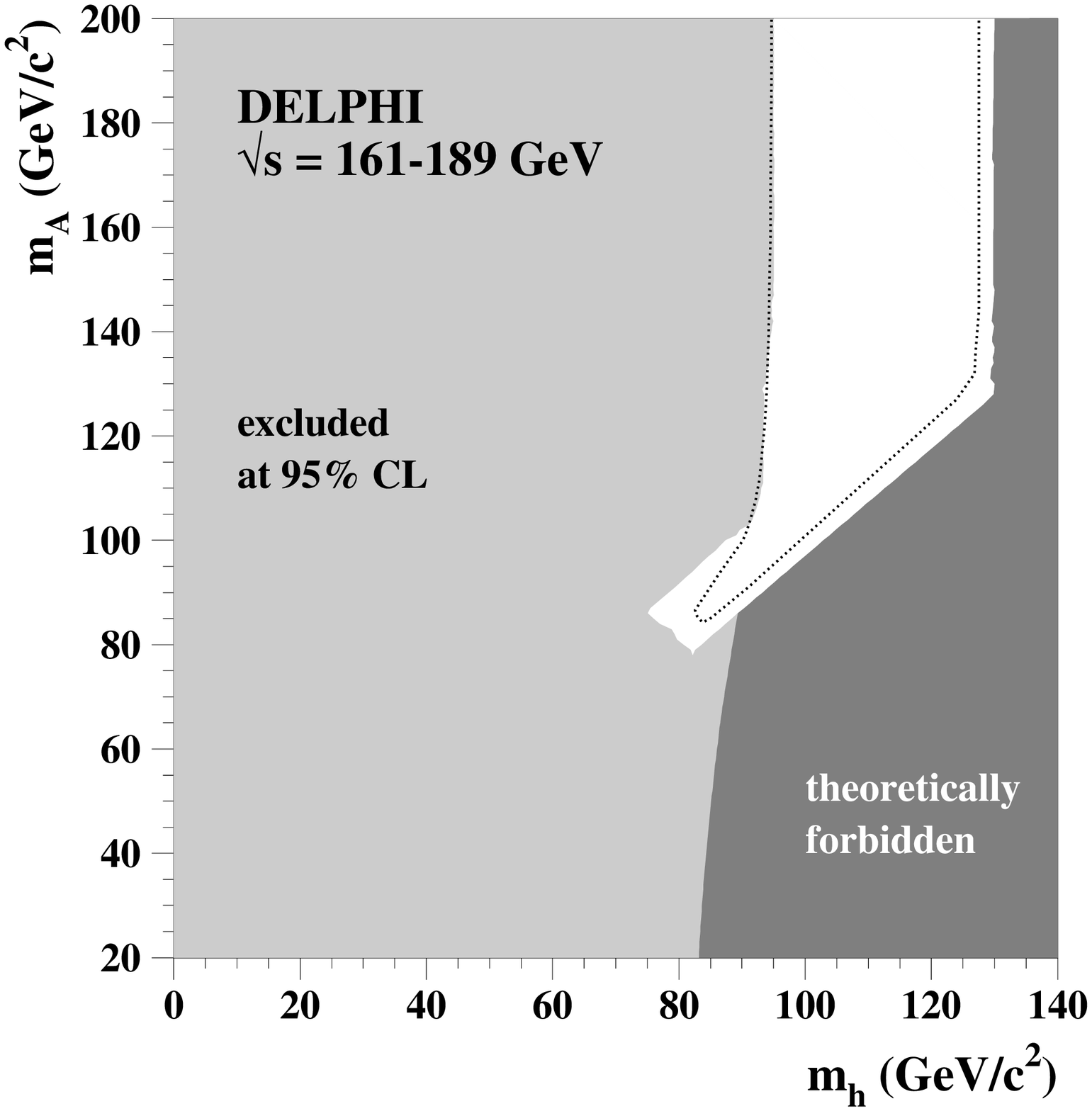}
\vspace*{-1.5cm}
\caption{189~GeV $(m_{\rm h},m_{\rm A})$ result. 
The dotted line indicates the
corresponding benchmark limit.}
\label{fig:189delphiha}
\vspace*{-3.5cm}
\end{figure}

The results from very different parameter scan methods
of different LEP experiments agree well as shown in
Figs.~\ref{fig:189opalha} and~\ref{fig:189delphiha}
for 189~GeV data~\cite{189bsopal,189s}.
The shaded region in Fig.~\ref{fig:189opalha} is
excluded by charge- and color-breaking (CCB) criteria~\cite{189bsopal}.

\pagebreak

\begin{figure}[h]
\vspace*{-1cm}
\includegraphics[width=0.47\textwidth]{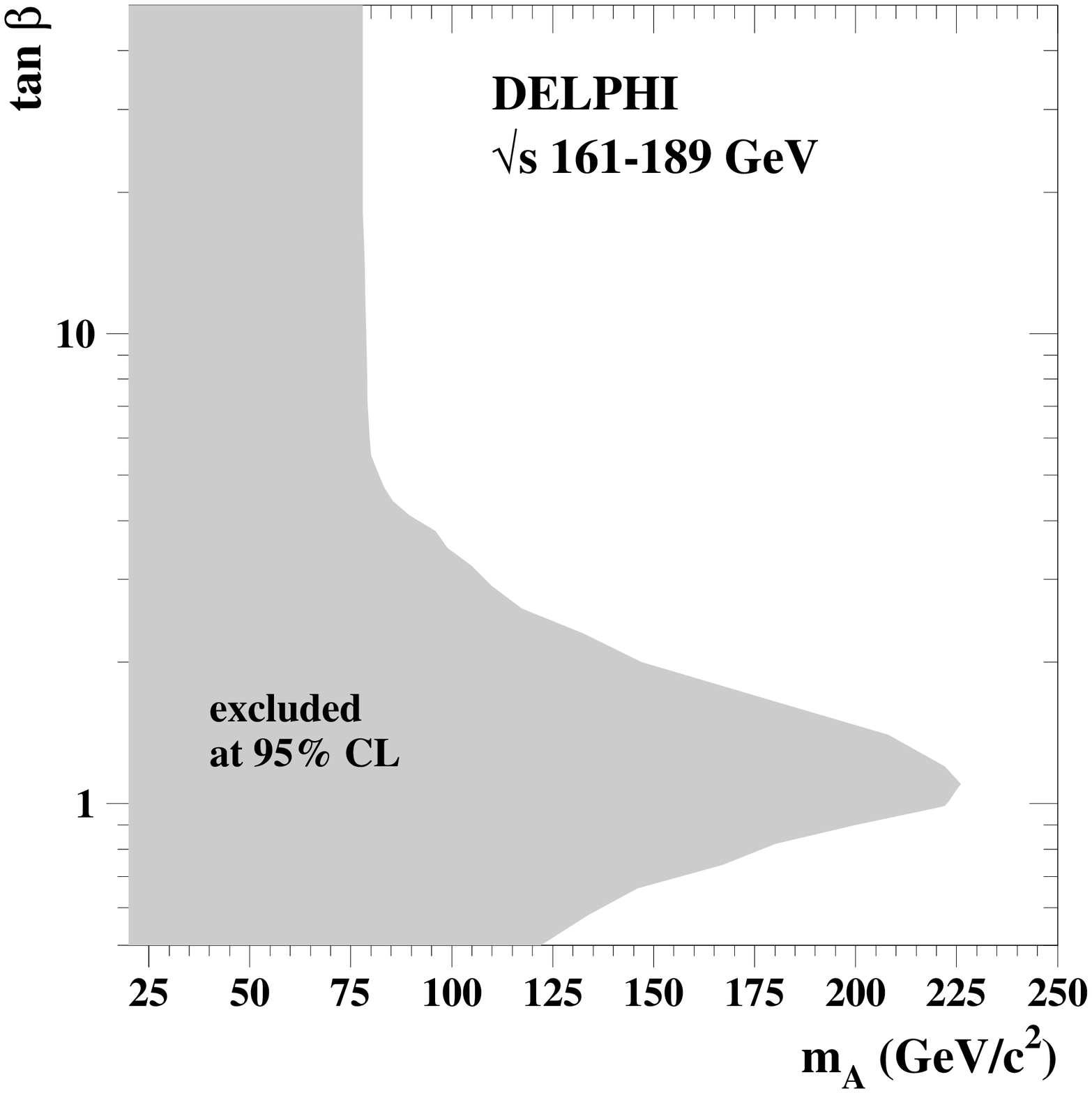}
\vspace*{-1.5cm}
\caption{189~GeV $(m_{\rm A},\tan\beta)$ result.}
\label{fig:189delphiatgb}
\includegraphics[width=0.47\textwidth]{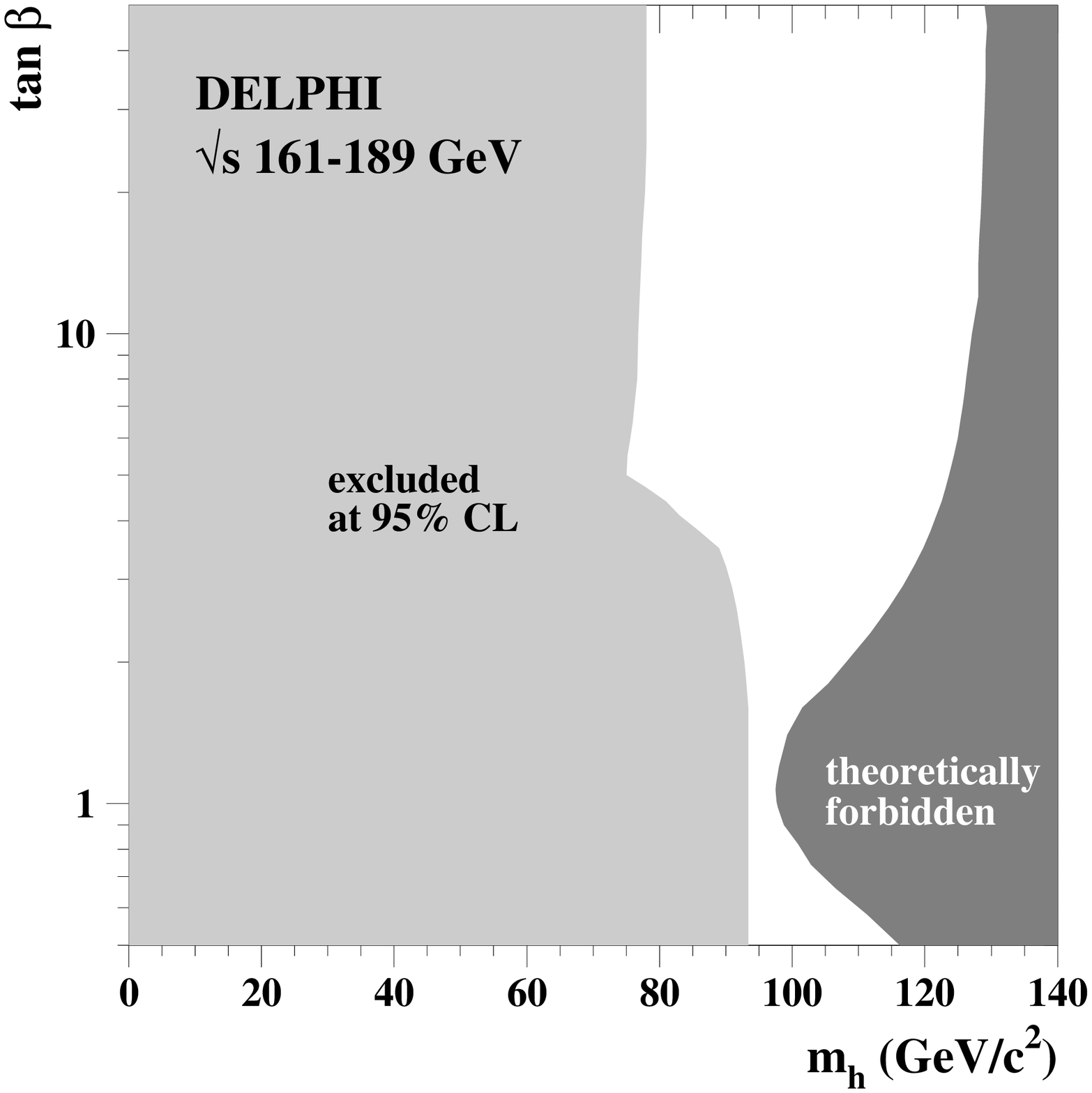}
\vspace*{-1.5cm}
\caption{189~GeV $(m_{\rm h},\tan\beta)$ result.}
\label{fig:189delphihtgb}
\vspace*{-1.5cm}
\end{figure}

Limits from charmless b-decays, the electroweak parameter $\Delta\rho$
and direct searches for Supersymmetric particles do not
change the h and A mass limits~\cite{189s}.
The parameter scan reduced the benchmark limits by
up to 7~GeV~\cite{189b,189s}.
The constraints on $\tan\beta$ are given in
Figs.~\ref{fig:189delphiatgb} and~\ref{fig:189delphihtgb}.

\clearpage
\section{Small reduction of limits}
\begin{figure}[h]
\vspace*{-1.2cm}
\includegraphics[width=0.5\textwidth]{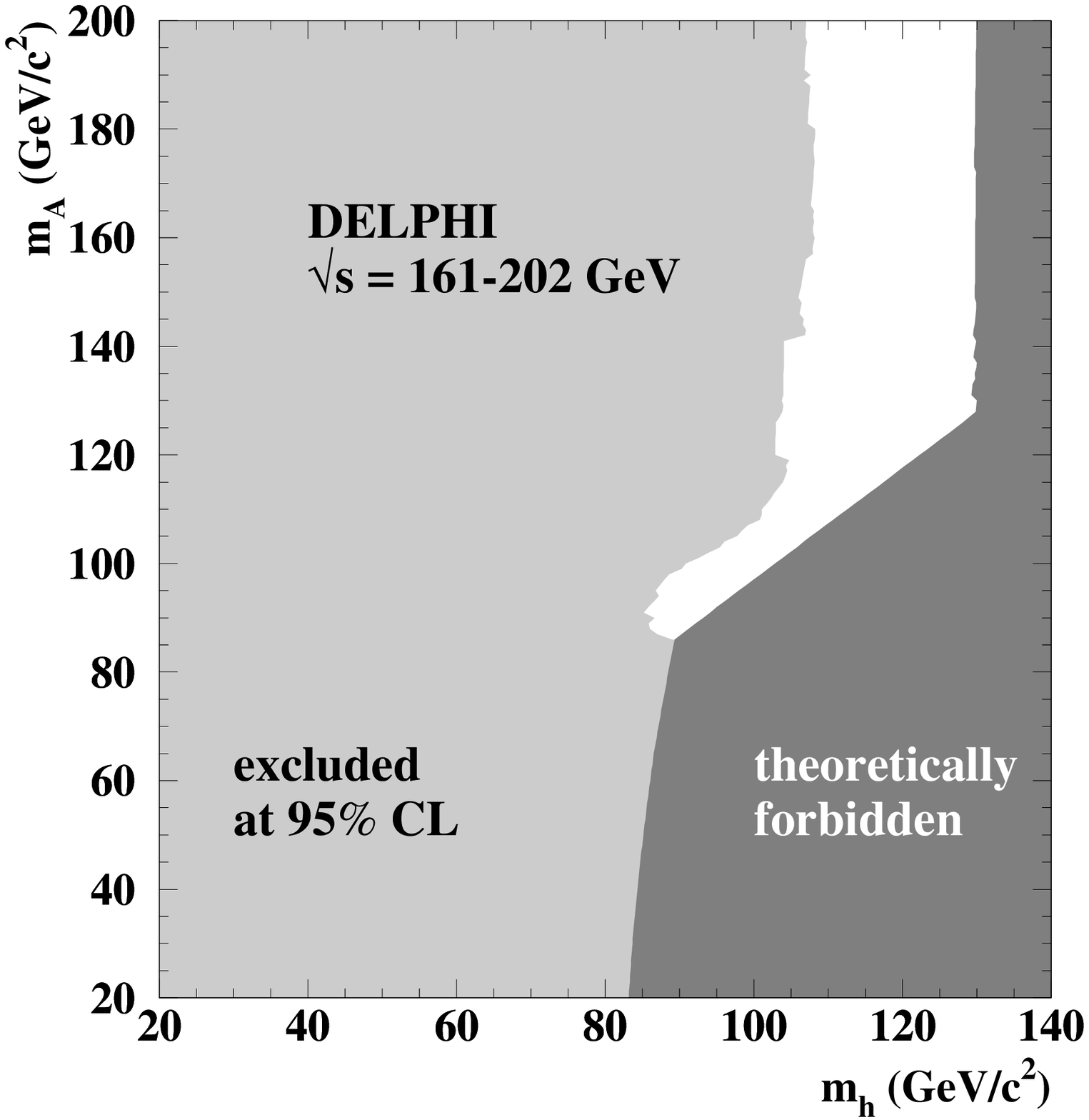}
\vspace*{-1.5cm}
\caption{202~GeV $(m_{\rm h},m_{\rm A})$ result.}
\label{fig:202delphiha}
\vspace*{-1.5cm}
\end{figure}

\begin{figure}[h]
\vspace*{-1cm}
\includegraphics[width=0.5\textwidth]{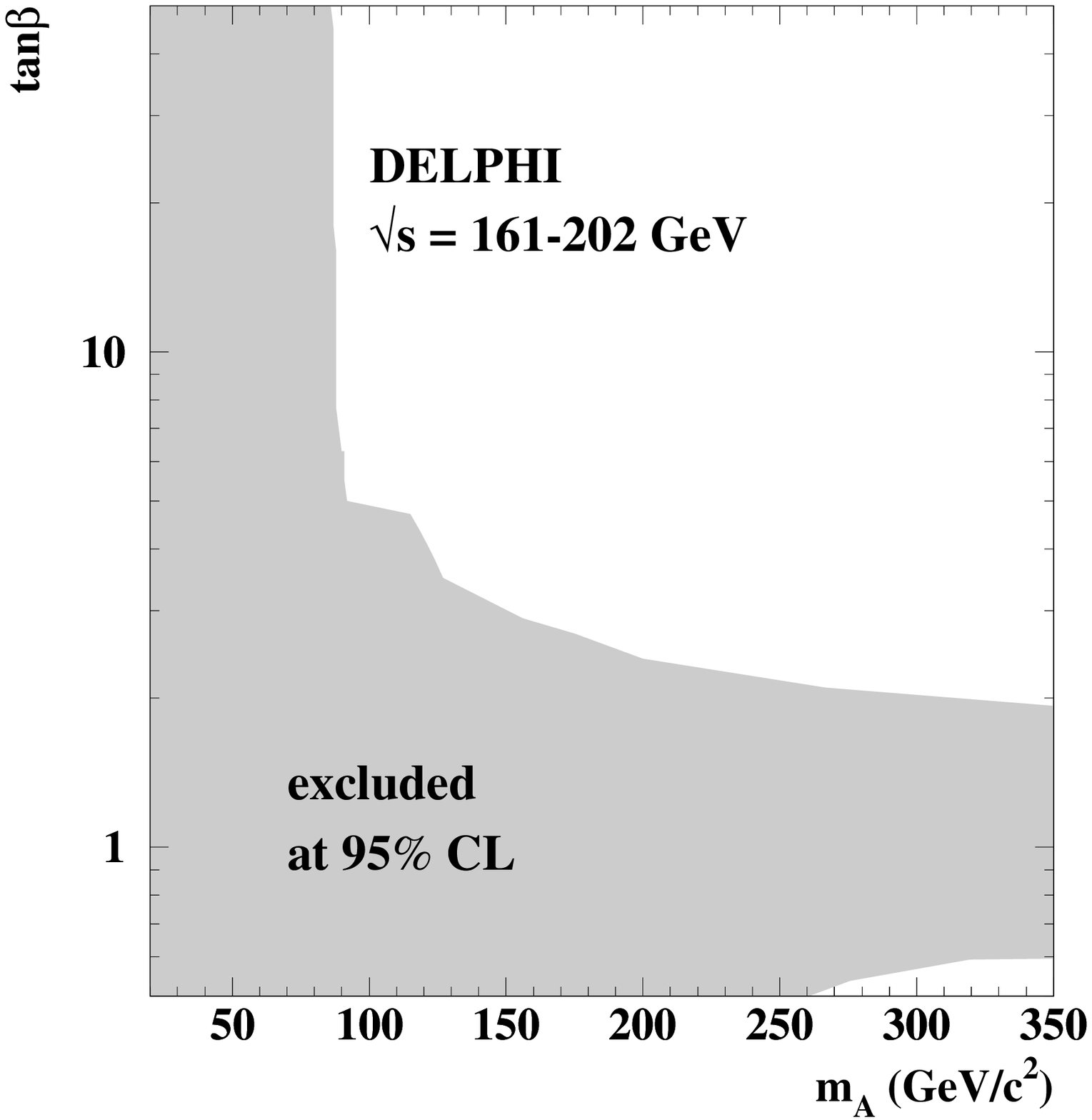}
\vspace*{-1.5cm}
\caption{202~GeV $(m_{\rm A},\tan\beta)$ result.}
\label{fig:202delphiatgb}
\vspace*{-1cm}
\end{figure}

\begin{figure}[t]
\vspace*{-0.7cm}
\includegraphics[width=0.5\textwidth]{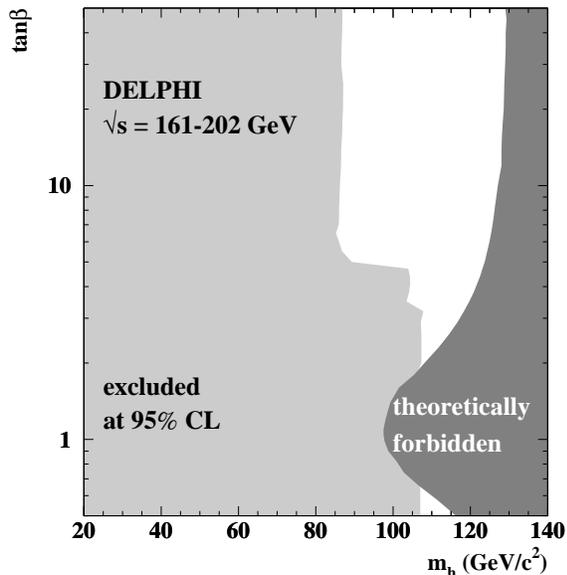}
\vspace*{-1.5cm}
\caption{202~GeV $(m_{\rm h},\tan\beta)$ result.}
\label{fig:202delphihtgb}
\vspace*{-0.2cm}
\end{figure}

With increasing center-of-mass energy, 
benchmark and parameter scan limits agree within 1~GeV
for 202~GeV data~\cite{202bs}.
The mass and $\tan\beta$ limits are given in 
Figs.~\ref{fig:202delphiha},~\ref{fig:202delphiatgb}
and~\ref{fig:202delphihtgb}.

\section{Combined LEP parameter scan limits}
\begin{figure}[h]
\vspace*{-0.5cm}
\includegraphics[width=0.47\textwidth]{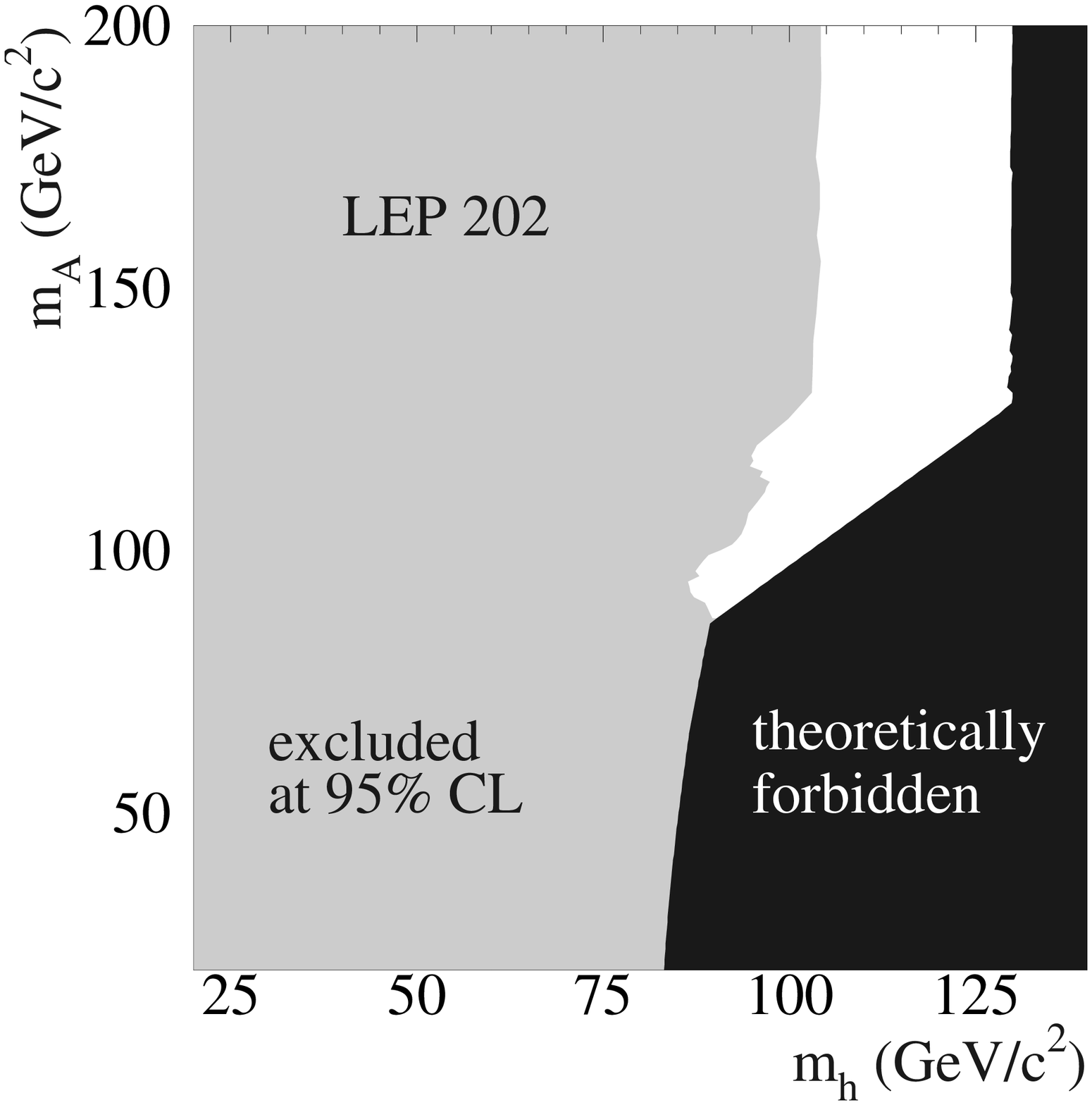}
\vspace*{-1.5cm}
\caption{202~GeV $(m_{\rm h},m_{\rm A})$ result.}
\label{fig:lep202delphiha}
\vspace*{-2cm}
\end{figure}

\begin{figure}[htb]
\includegraphics[width=0.47\textwidth]{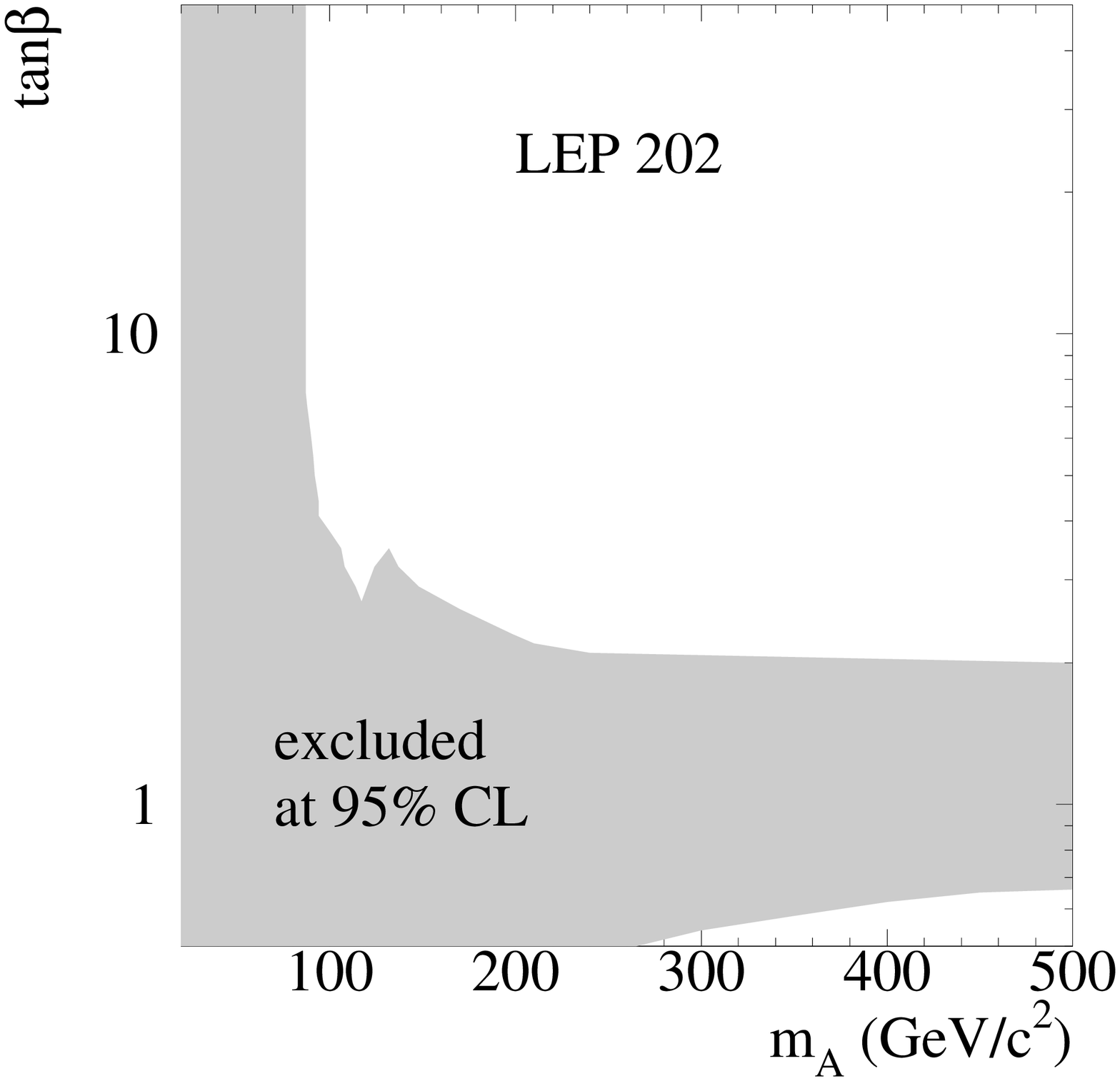}
\vspace*{-1.5cm}
\caption{202~GeV $(m_{\rm A},\tan\beta)$ result.}
\label{fig:lep202delphiatgb}
\vspace*{-1cm}
\end{figure}

\begin{figure}[htb]
\vspace*{-0.7cm}
\includegraphics[width=0.47\textwidth]{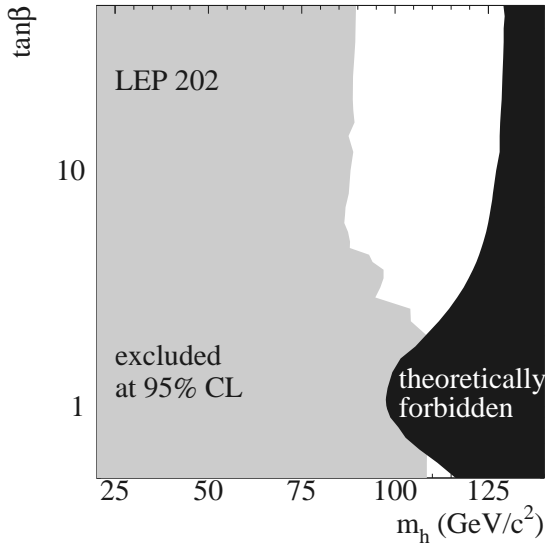}
\vspace*{-1.5cm}
\caption{202~GeV $(m_{\rm h},\tan\beta)$ result.}
\label{fig:lep202delphihtgb}
\vspace*{-1cm}
\end{figure}

The mass limits from benchmark and parameter scan
agree within about 2~GeV in the
combination of the 202~GeV data from all LEP 
experiments~\cite{202ball,202sall}.
The scan limits are given in
Figs.~\ref{fig:lep202delphiha},~\ref{fig:lep202delphiatgb}
and~\ref{fig:lep202delphihtgb}.

\clearpage

\section{Latest results}
\begin{figure}[htb]
\vspace*{-0.9cm}
\includegraphics[width=0.47\textwidth]{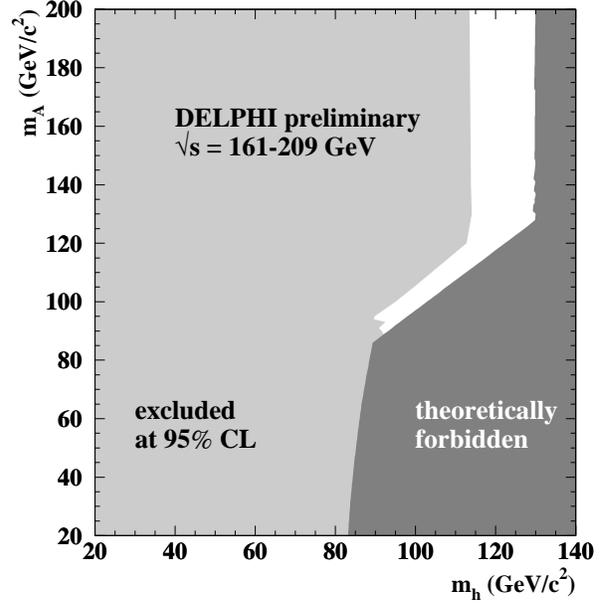}
\vspace*{-1.3cm}
\caption{209~GeV $(m_{\rm h},m_{\rm A})$ result.}
\label{fig:209delphiha}
\vspace*{-0.8cm}
\end{figure}

\begin{figure}[htb]
\vspace*{-1cm}
\includegraphics[width=0.47\textwidth]{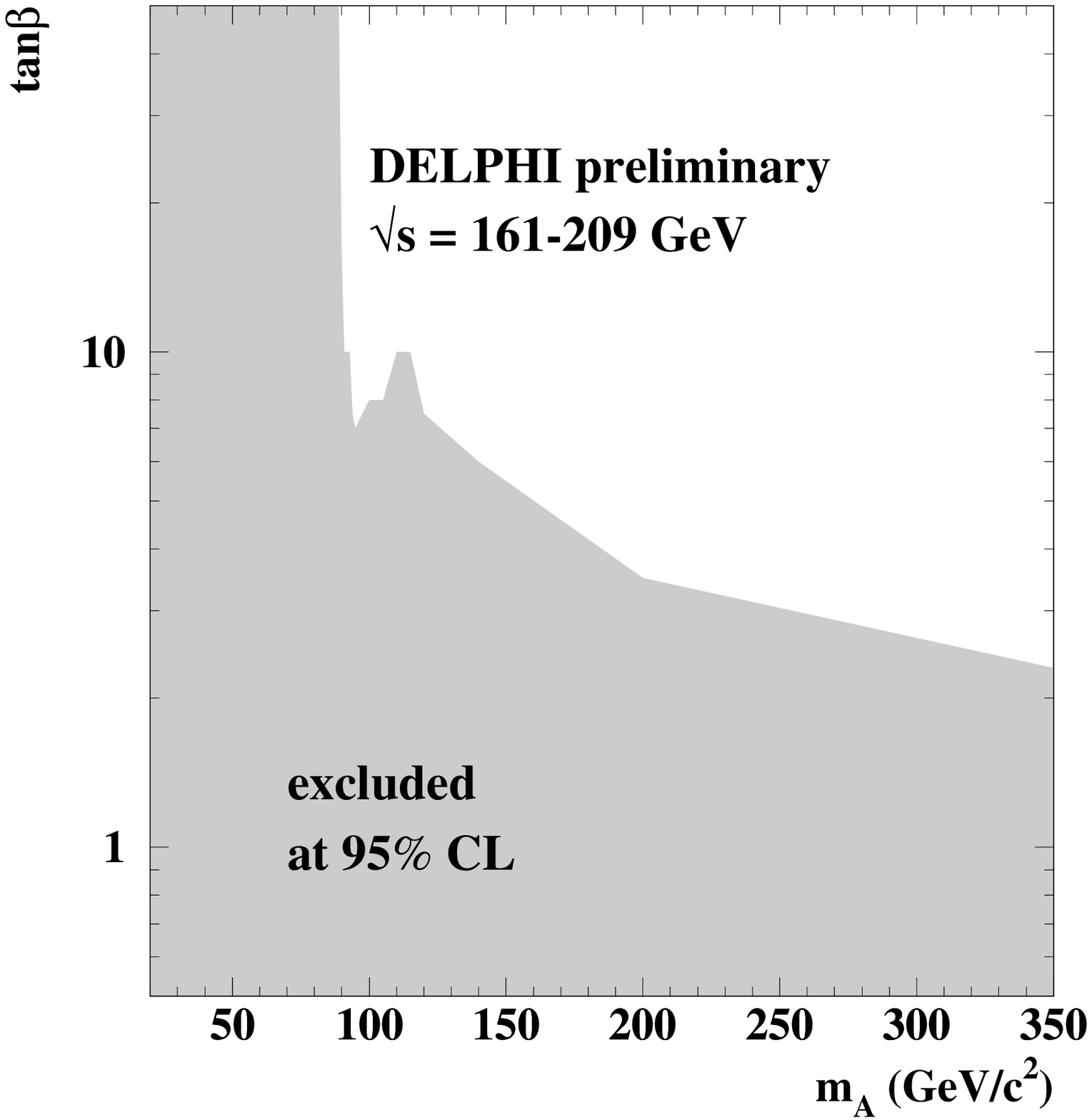}
\vspace*{-1.3cm}
\caption{209~GeV $(m_{\rm A},\tan\beta)$ result.}
\label{fig:209delphiatgb}
\vspace*{-3.4cm}
\end{figure}

\begin{figure}[htb]
\vspace*{-1cm}
\includegraphics[width=0.47\textwidth]{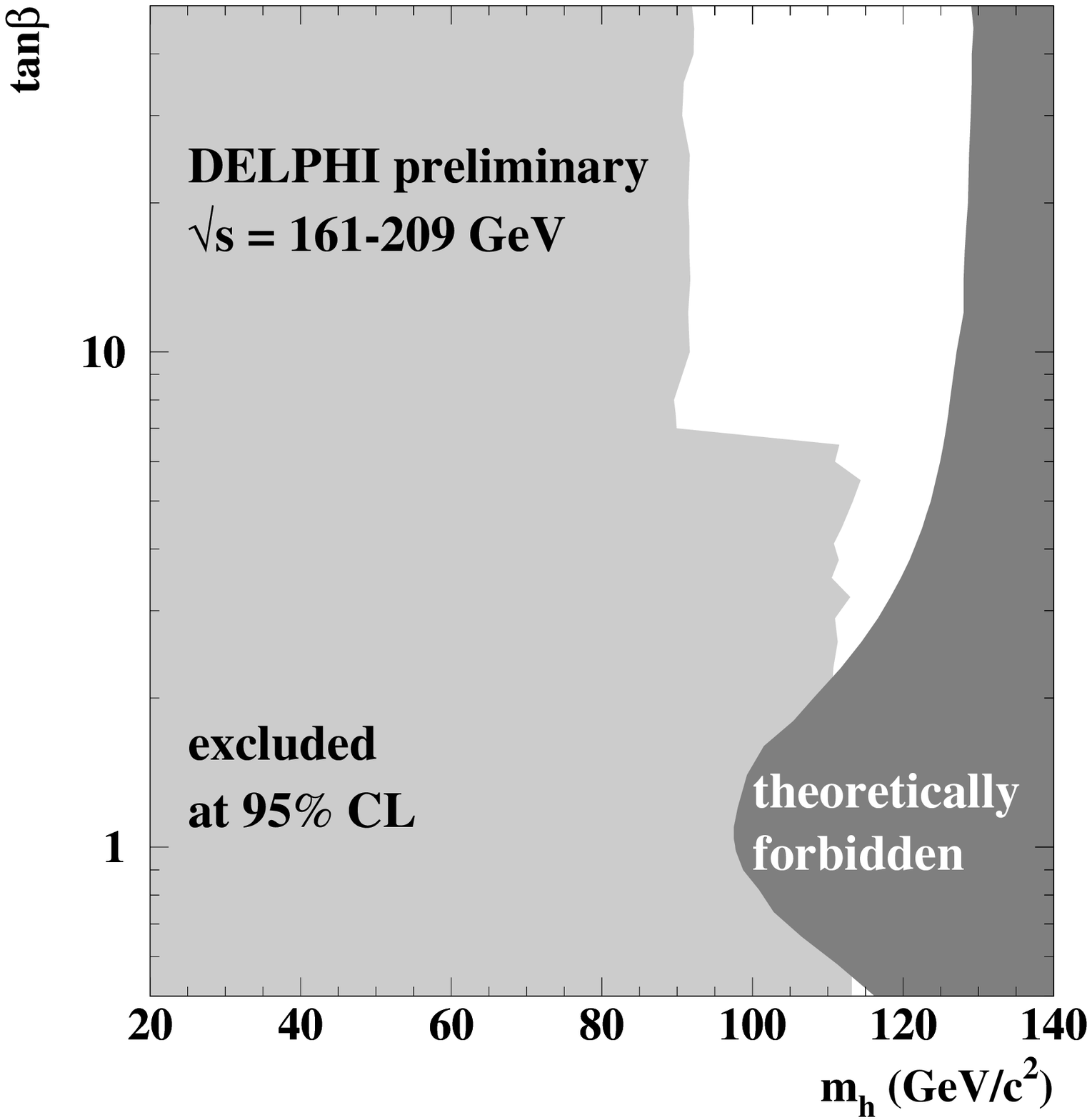}
\vspace*{-1.5cm}
\caption{209~GeV $(m_{\rm h},\tan\beta)$ result.}
\label{fig:209delphihtgb}
\vspace*{-0.3cm}
\end{figure}

\begin{figure}[htb]
\vspace*{-0.8cm}
\includegraphics[width=0.47\textwidth]{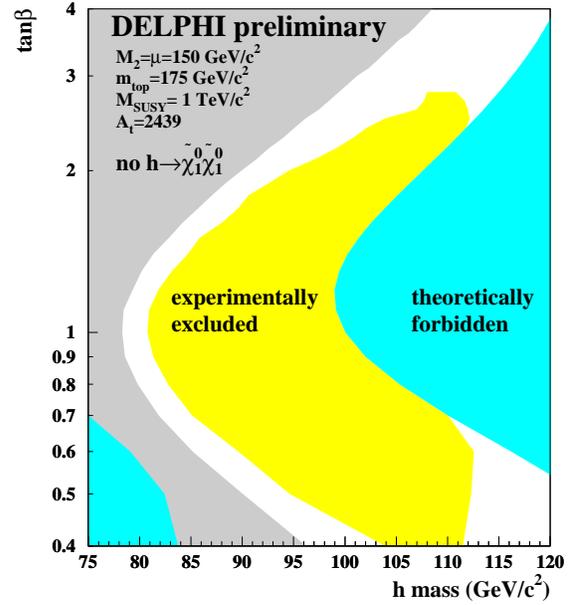}
\vspace*{-1.1cm}
\caption{209~GeV $(m_{\rm h},\tan\beta)$ result from searches for
         invisibly decaying Higgs bosons.}
\label{fig:209inv}
\vspace*{-1cm}
\end{figure}

The latest benchmark and scan results including 209~GeV data
agree within 2~GeV~\cite{209b,209s} and 
give mass limits on h and A of 89~GeV
(Figs.~\ref{fig:209delphiha},~\ref{fig:209delphiatgb}
and~\ref{fig:209delphihtgb}).
The importance of a MSSM parameter scan is underlined in
Fig.~\ref{fig:209inv}, which shows that large parameter
regions exist where a Higgs boson could decay invisibly into 
a pair of neutralinos. Thus, only in conjunction with 
a dedicated search~\cite{delphiinv} the h and A mass limits
are set.

\clearpage

\section{Three Higgs boson hypothesis}
\begin{figure}[htb]
\begin{center}
\vspace*{0.8cm}
\includegraphics[width=0.47\textwidth]{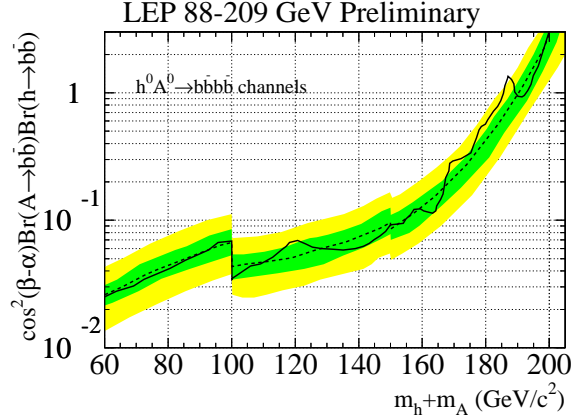}
\vspace*{-1.7cm}
\caption[]{
Limits on the hA cross section as a function
of $\mh+\mA$ at 95\% CL ($\mh\approx\mA$) for the MSSM processes
\ee\ra~hA~\ra~\bb\bb.
This corresponds also to limits on $\cos^2(\beta-\alpha)$ in the
general extension of the SM with two Higgs boson doublets.
The data of the four LEP experiments
collected at energies from 88 to 209~GeV are combined.  
The solid curve is the observed result and the dashed curve shows the 
expected median.
Shaded areas indicate the $1\sigma$ and $2\sigma$ probability bands.
\label{mssm-bbbb}}
\end{center}
\vspace*{-0.4cm}
\end{figure}

\begin{figure}[t]
\begin{center}
\vspace*{-2.5cm}
\includegraphics[width=0.47\textwidth]{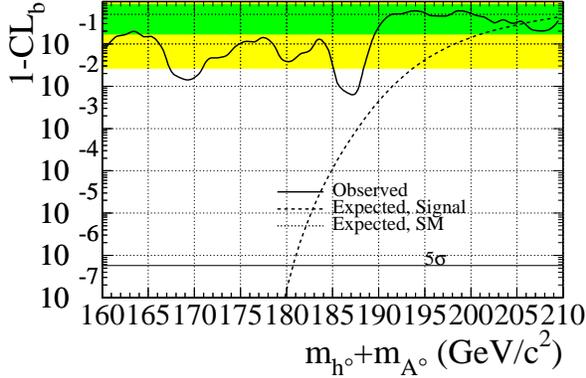}
\vspace*{-1.7cm}
\caption[]{
The confidence level $1-CL_{\rm b}$ 
as a function of $\mh+\mA$ for the case $\mh\approx\mA$
(where only the \ee\ra~hA process contributes since $\sba\approx0$). 
The straight line at 0.5 and the shaded $1\sigma$ and $2\sigma$ probability 
bands represent the expected background-only result.
The solid curve is the observed result and the dashed curve shows the 
expected median for a signal.
\label{mssm-diag}}
\end{center}
\vspace*{-2cm}
\end{figure}

\begin{figure}[b]
\vspace*{-1cm}
\begin{center}
\includegraphics[width=0.47\textwidth]{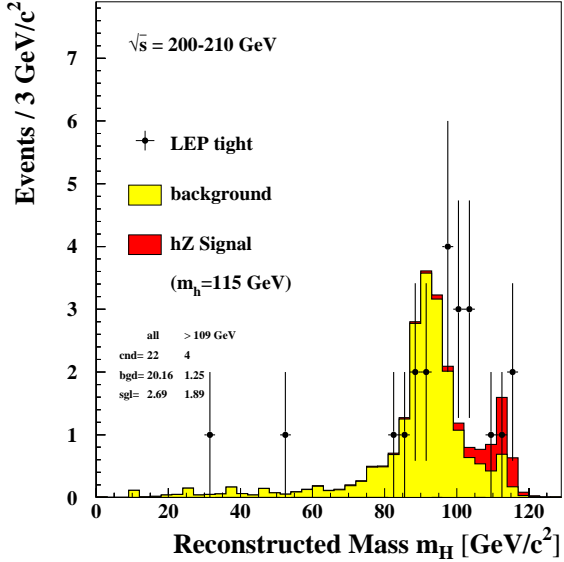}
\vspace*{-1.7cm}
\caption[]{
Distribution of the reconstructed SM Higgs boson mass in searches 
conducted at energies between 200 and 210 GeV.
The figure displays the data
(dots with error bars), the predicted SM background
and the prediction for a Higgs boson of 115~GeV mass.
The number of data events selected with mass larger than 109 GeV
is 4, while 1.25 are expected from SM background processes and
1.89 from a 115 GeV signal.
Between 96 and 105~GeV 10 data events are observed, while 3.6 
background events are expected.
\label{sm-mass}}
\end{center}
\vspace*{-2cm}
\end{figure}

\begin{figure}[t]
\vspace*{-1.7cm}
\begin{center}
\includegraphics[width=0.47\textwidth]{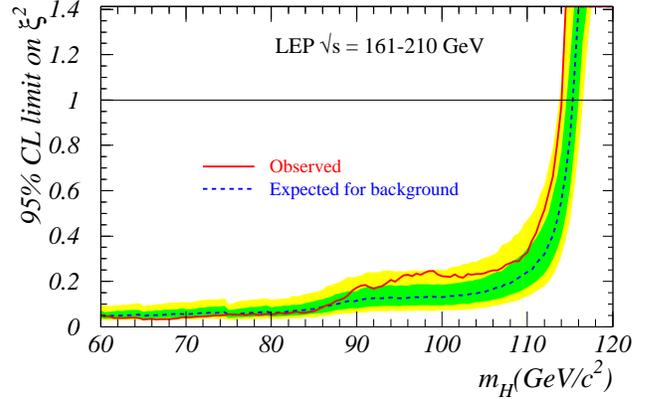}
\vspace*{-1.3cm}
\caption[]{
The 95\% CL upper bound on $\xi^2$ as a function 
of \mH, where $\xi= g_{\rm HZZ}/g_{\rm HZZ}^{\rm SM}$ is the HZZ coupling
relative to the SM coupling. About $2\sigma$ deviations from the
expectation are observed at $m_{\rm H}=98$~GeV and $m_{\rm H}=115$~GeV.
In the MSSM, hZ production at the lower mass and HZ production at the
higher mass are possible. 
\label{sm-xi-1}}
\end{center}
\vspace*{-0.3cm}
\end{figure}

\begin{figure}[b]
\vspace*{-2cm}
\begin{center}
\includegraphics[width=0.47\textwidth]{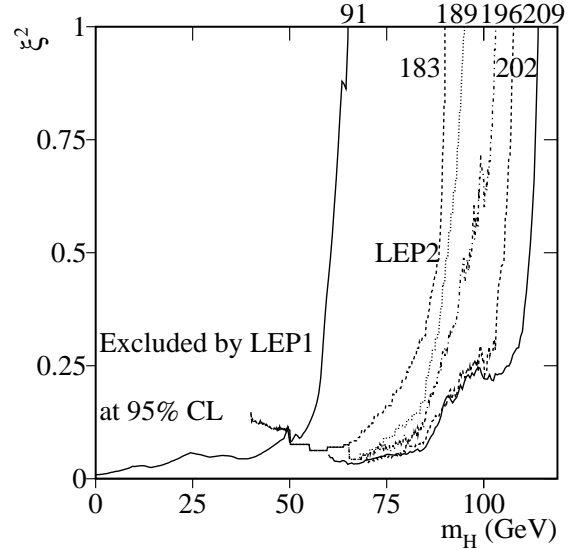}
\vspace*{-1.3cm}
\caption[]{
The excluded $(\xi^2,m_{\rm H})$ region including 209 GeV data is 
compared with the results
from combined LEP-1 data~\cite{lep1}, taken around 91 GeV center-of-mass 
energy, and previous LEP-2 limits~\cite{202sall} up to 
183, 189, 196 and 202 GeV. 
The $\xi^2$ limit below 100~GeV does not become significantly stronger
when the 209 GeV data, taken in 2000, are included. 
\label{sm-xi-2}}
\end{center}
\vspace*{-2cm}
\end{figure}

For 202 GeV data, taken in 1999,
and first 2000 data, h and A mass limits were 2 GeV below
expectation~\cite{as2000}.
This tendency is enhanced by including complete 209 GeV
data, in which case the limits are 3.1 to 3.6 GeV 
below the expectations of about 95~GeV~\cite{mssmlep209}.
A possible explanation is that the HZ excess at about 115~GeV is due 
to the heavier scalar H and that, in addition, the production of hA with 
masses between 90 and 100 GeV occurs~\cite{as2000}.
Figure~\ref{mssm-bbbb} (from~\cite{mssmlep209}) shows a data excess above 
$2\sigma$ for $m_{\rm h}+m_{\rm A}=187$~GeV in the \bb\bb\ channel.
The same data excess is also expressed by the confidence level
$CL_{\rm b}$ for a signal observation
as shown in Fig.~\ref{mssm-diag} (from~\cite{mssmlep209}).
The hypothesis of the production of three MSSM Higgs bosons is supported
by the data excess seen in Fig.~\ref{sm-mass} (from~\cite{smlep209})
at 100 GeV which could result from hZ production in addition to HZ production.
For the reported MSSM parameters~\cite{as2000}
$\cos^2(\beta-\alpha)\approx 0.9$; therefore 
$\sin^2(\beta-\alpha)=\xi^2\approx0.1$.
The $\xi^2$ limit in the 100 GeV mass region shows a deviation of about
2$\sigma$ between the expected and observed limit,
as seen in Fig.~\ref{sm-xi-1} (from~\cite{smlep209}).
Figure~\ref{sm-xi-2} shows that this new support is only observed
in the complete LEP data.

\clearpage

\clearpage
\section{Conclusions}
\vspace*{-0.1cm}
The LEP-1 and LEP-2 data are consistent with a background-only hypothesis 
and give stringent mass limits on the neutral Higgs bosons h and A of 
the MSSM.
During the LEP era, MSSM parameter scans reduced -- 
or for early LEP-2 data even removed completely -- the benchmark mass limits.
The importance of parameter scans is stressed by large parameter regions
where the Higgs boson decays invisibly, which is not considered 
in benchmark results. 
At the highest center-of-mass energies,
benchmark limits are only slightly reduced by a general parameter
scan when the results from invisible Higgs boson searches are included.
Table~\ref{tab:comp} compares benchmark and scan mass limits 
in the MSSM.

The combined LEP data show a preference for the 
SM Higgs boson of 115.6~GeV, which can also be 
interpreted as a preference for a Higgs boson of
that mass in the MSSM. Further small data excesses
for Higgs boson pair-production and brems\-strahlung
between 90 and 100~GeV allow the hypo\-thesis 
that h, A and H of the MSSM all have 
masses below 116~GeV. Previously reported MSSM 
parameter combinations from a general parameter scan 
for this scenario are supported by the complete data set.

\section*{Acknowledgments}
\vspace*{-0.1cm}
I would like to thank the organizers of the conference
for their kind hospitality.

\begin{table}[htbp]
\vspace*{-0.5cm}
\caption{Benchmark (b) and scan (s) mass limits in the MSSM.
All limits are in GeV at 95\% CL.}
\label{tab:comp}
\renewcommand{\arraystretch}{1.2} 
\begin{tabular}{ll|rr|rr}
$\sqrt{s}$ (GeV)&Data& $m_{\rm h}^{\rm b}$ &$m_{\rm A}^{\rm b}$ &$m_{\rm h}^{\rm s}$ &$m_{\rm A}^{\rm s}$ \\\hline
91~\cite{91b,91s}   & L3      & 41.0 & none      & 25  & none \\
172~\cite{172b,172s}& DELPHI  & 59.5 & 51.0      & 30  & none \\
183~\cite{183b,183s}& DELPHI  & 74.4 & 75.2      & 67  & 75   \\
189~\cite{189bsopal}& OPAL&74.8& 76.5            & 72  & 76  \\
189~\cite{189b,189s}& DELPHI  & 82.6 & 84.1      & 75  & 78   \\
202~\cite{202bs}& DELPHI  & 85.9 & 86.5      & 85  & 86   \\
202~\cite{202ball,202sall}& LEP&88.3& 88.4      & 86  & 87   \\
209~\cite{209b,209s}& DELPHI  & 89.6 & 90.7      & 89  & 89   
\end{tabular}
\end{table}

\end{document}